\documentclass[%
 aip,
 apl,
 amsmath,amssymb,
 reprint,%
]{revtex4-1}

\usepackage[utf8]{inputenc}
\usepackage{color}
\usepackage{soul}
\usepackage[alsoload=synchem]{siunitx}
\usepackage[version=3]{mhchem}
\usepackage{graphicx}
\usepackage{dcolumn}
\usepackage{bm}
\usepackage[acronym,shortcuts]{glossaries}
\usepackage[percent]{overpic}
\usepackage{url}
\usepackage[breaklinks,colorlinks=true,allcolors=blue]{hyperref}
\usepackage{subfigure}
\usepackage{gensymb}

\begin{document}

\preprint{AIP/123-QED}

\title{Real-time buffer gas pressure tuning in a micro-machined vapor cell}

\author{S. Dyer}
\affiliation{Department of Physics, SUPA, University of Strathclyde, Glasgow G4 0NG, United Kingdom}
\author{A. McWilliam}
\affiliation{Department of Physics, SUPA, University of Strathclyde, Glasgow G4 0NG, United Kingdom}
\author{D. Hunter}
\affiliation{Department of Physics, SUPA, University of Strathclyde, Glasgow G4 0NG, United Kingdom}
\author{S. Ingleby}
\affiliation{Department of Physics, SUPA, University of Strathclyde, Glasgow G4 0NG, United Kingdom}
\author{D. P. Burt}
\affiliation{Kelvin Nanotechnology, University of Glasgow, Glasgow  G12 8LS, United Kingdom}
\author{O. Sharp}
\affiliation{Kelvin Nanotechnology, University of Glasgow, Glasgow  G12 8LS, United Kingdom}
\author{F. Mirando}
\affiliation{Kelvin Nanotechnology, University of Glasgow, Glasgow  G12 8LS, United Kingdom}
\author{P. F. Griffin}
\affiliation{Department of Physics, SUPA, University of Strathclyde, Glasgow G4 0NG, United Kingdom}
\author{E. Riis}
\affiliation{Department of Physics, SUPA, University of Strathclyde, Glasgow G4 0NG, United Kingdom}
\author{J. P. McGilligan}
\affiliation{Department of Physics, SUPA, University of Strathclyde, Glasgow G4 0NG, United Kingdom}
\email{james.mcgilligan@strath.ac.uk}

\date{\today}

\begin{abstract}
We demonstrate a controllable depletion of the nitrogen buffer gas pressure in a micro-machined cesium (Cs) vapor cell from the dynamic heating of an alkali dispenser pill. When the alkali source is laser activated, the gettering compounds within the alkali pill dispenser reduce the nitrogen (N$_2$) content from the vapor for fine-tuning of the alkali to buffer gas pressure ratio. Additionally, we decrease the buffer gas pressure below 100$\,$mTorr to evaluate the presence of other potential broadening mechanisms. Real-time control of the gas pressure ratio in the vapor cell will have notable benefits for refining atomic sensor performance and provide a routine to achieve various target pressures across a wafer bonded with a uniform back-filled buffer gas pressure.

\end{abstract}

\maketitle

Precision spectroscopy of atomic ensembles remains at the forefront of quantum technologies and state-of-the-art metrological instruments. A prominent platform used to this end is the interrogation of atoms in the vapor phase, which provides a relatively simple apparatus for probing and controlling the atomic state. The implementation of vapor phase atomic platforms range from investigations of fundamental physics \cite{ackemann2000,Offer2018}, to applied research in magnetometry \cite{Budker2007}, electrometry \cite{natphyselectro}, and terahertz sensing \cite{KevTerahertz} among other applications \cite{Kitching2018, moshe}.

In recent years, the miniaturization of atom-optic technology has advanced the state of field-deployable sensors for practical use in real-world applications \cite{Kitching2018,mcgilliganreview}. At the heart of many of these instruments is a micro-electro-mechanical-systems (MEMS) alkali vapor cell, that provides a stage for atom-light interactions \cite{moreland}. The MEMS cells typically consist of an etched or mechanically cut silicon cavity sandwiched between two anodically bonded glass wafers. The design versatility and mass-producible fabrication processes have resulted in MEMS cells being widely implemented in compact clocks \cite{Knappe2004, batori, Carle:23, Gusching:23}, magnetometers\cite{microfabcoilkitching,Hunter:22,Ingleby2022}, and wavelength references \cite{Hummon:18,Hafiz:16} as well as recently transferring to cold-atom technology \cite{dyer2023PRApplied, McGilligan2020, Bregazzi}. 

A key step in the fabrication of the MEMS cell is the deposition technique used for the introduction of alkali atoms into the vapor cell. While the purity of the atomic species within the cell is critical to decrease unwanted collisional shifts and broadening, the ratio of the alkali and buffer gas plays an essential role in the performance of the atomic sensor \cite{HASEGAWA2011594}. A number of options have been demonstrated as mass-producible solutions to vapor deposition, such as alkali-azides\cite{Terry2022,morelandazide}, chlorides\cite{Bopp_2020,Knappechloride}, graphite reservoirs \cite{kang-roper} and chromate based micro-pill dispensers \cite{boudotpill,HASEGAWA2011594}. For each of these methods, the inclusion of a substantial additional buffer gas pressure is typically achieved by back-filling the vacuum bonder with the desired mixture of gases prior to cell encapsulation \cite{Kitching2018}. However, this approach can result in cell-to-cell variations across the wafer, as has been recently shown in the case of alkali-azide deposition \cite{Terry2022}. While deviations from the target pressure can be addressed by changing the operational temperature of the cell \cite{Kroemer:15, HASEGAWA2011594}, this approach would be unfavourable for integrated optics packages were the cell and laser require stabilization at a common value \cite{boudotmemscell}. Importantly, the method of back-filling a buffer gas pressure during bonding is applied across an entire wafer, such that all cells derived from the same wafer are restricted to a single target buffer gas pressure and composition, with additional cell complexity required for individual cell tunability \cite{Maurice2022}.

In this Letter, we demonstrate the real-time fine tuning of the nitrogen buffer gas pressure within a micro-machined cesium vapor cell. A thick silicon wafer is bonded to borosilicate glass within a nitrogen environment, with an alkali pill deposited within the silicon frame. When laser activated, the pill source provides a saturated cesium vapor pressure within each cell. We show that the initial back-filled nitrogen pressure can be controllably decrease by laser heating the non-evaporable getter compound within the dispensing pill. In this work we show by tracking the broadening and pressure shift in the sub-Doppler features of $D_1$ Cs spectroscopy during pill heating, that the nitrogen content can be fine-tuned over a pressure range of 150$\,$Torr to below 100$\,$mTorr. The stability of the target pressure within the cell is shown to remain stable over a prolonged period of months after the heating of the pill has stopped. Finally, the zero-intensity linewidth is measured for a cell with a greatly diminished buffer gas pressure. This approach could prove beneficial to the fine-tuning of buffer gas cells in a plethora of atomic sensors, while also enabling fabrication routines that realize many individual target buffer gas pressures derived from the constituent cells of a single wafer.
\begin{figure}[t]
\centering
\includegraphics[width=\columnwidth]{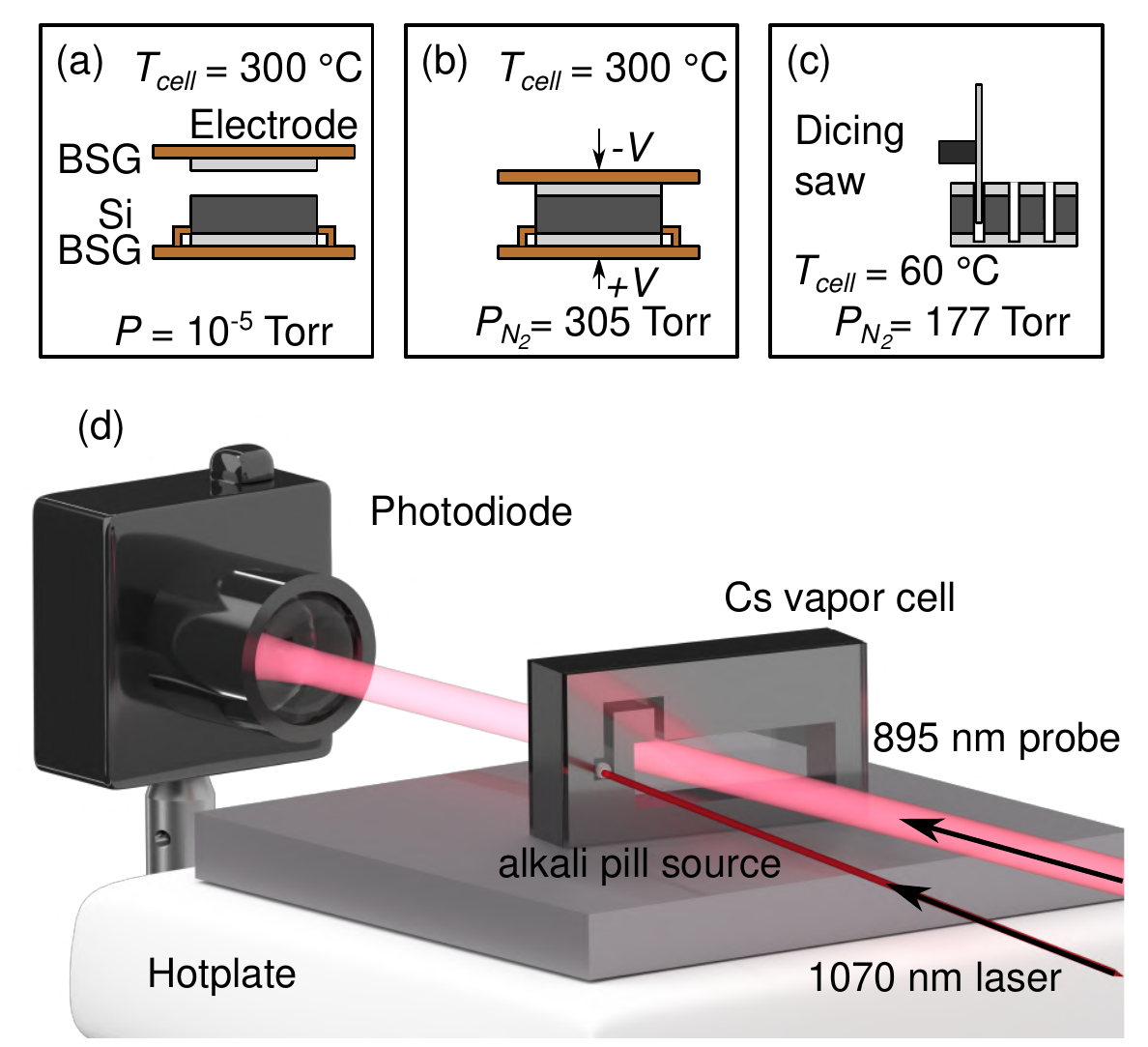}
\caption{(a)-(c): Wafer bonding procedure. Wafers are heated to 300$\,^{\circ}$C and chamber is pumped down to a vacuum pressure of $P=10^{-5}\,$Torr. Nitrogen is back-filled to the target pressure at this temperature. The wafer is then diced into constituent vapor cells and measured at $T_{cell}=60\,^{\circ}$C. (d): Illustration of the measurement apparatus. The Cs vapor cell is heated on a hotplate while absorption of the 894.6$\,$nm probe is monitored. Activation is performed with a 1070$\,$nm laser aligned to an alkali pill dispenser within the micro-machined vapor cell.}
\label{bonding}
\end{figure}

A simplified schematic of the vapor cell bonding and measurement procedure is shown in Fig.~\ref{bonding}. A 6$\,$mm thick silicon wafer is water-jet cut to the desired cavity geometry, with two chambers connected via a meandering channel to avoid glass darkening from pill activation in the spectroscopy region of the cell \cite{dyer2022}. The processed silicon wafer is first pre-bonded to a lower borosilicate glass (BSG) wafer. The pill dispensers are then deposited within the silicon frame. The final bond is carried out within a vacuum apparatus, with the process illustrated in Fig.~\ref{bonding} (a). The pre-formed wafer stack and upper BSG wafer are held on independent surface electrodes while the vacuum is pumped down to the order of 10$^{-5}\,$Torr. Following vacuum evacuation, nitrogen is back-filled within the chamber to a pressure of 305$\,$Torr while the wafers are heated to 300$\,^{\circ}$C, shown in Fig.~\ref{bonding} (b). A high voltage is applied across the wafers to anodically bond a hermetic seal for vacuum encapsulation. The wafers are then removed from the chamber and mechanically diced into constituent vapor cells, shown in Fig.~\ref{bonding} (c). An individual cell is then placed within a spectroscopy set-up and heated to an operational temperature of 60$\,^{\circ}$C on a hotplate, illustrated in Fig.~\ref{bonding} (d).

An 894.6$\,$nm distributed Bragg reflector (DBR) laser is used to resolve spectroscopy of the $D_1$ transitions from the $F=3$ and $F=4$ ground-states of Cs. The light is fibre coupled and split into 3 arms. The first arm is coupled into a Fabry-P\'erot cavity to calibrate the frequency scan from the laser. The second arm is aligned through a 7-cm-long Cs vapor cell to calibrate the frequency scan with respect to the Cs $D_1$ spectrum. The third arm is aligned through the main chamber of the MEMS cell onto a photodiode to record absorption spectra for each buffer gas pressure during the tuning process. The probe beam has a beam waist of 700$\,\mu$m (1/e$^2$ radius) and average intensity of 75$\,\mu$W/cm$^2$. 
\begin{figure}[t!]
\centering
\includegraphics[width=1\columnwidth]{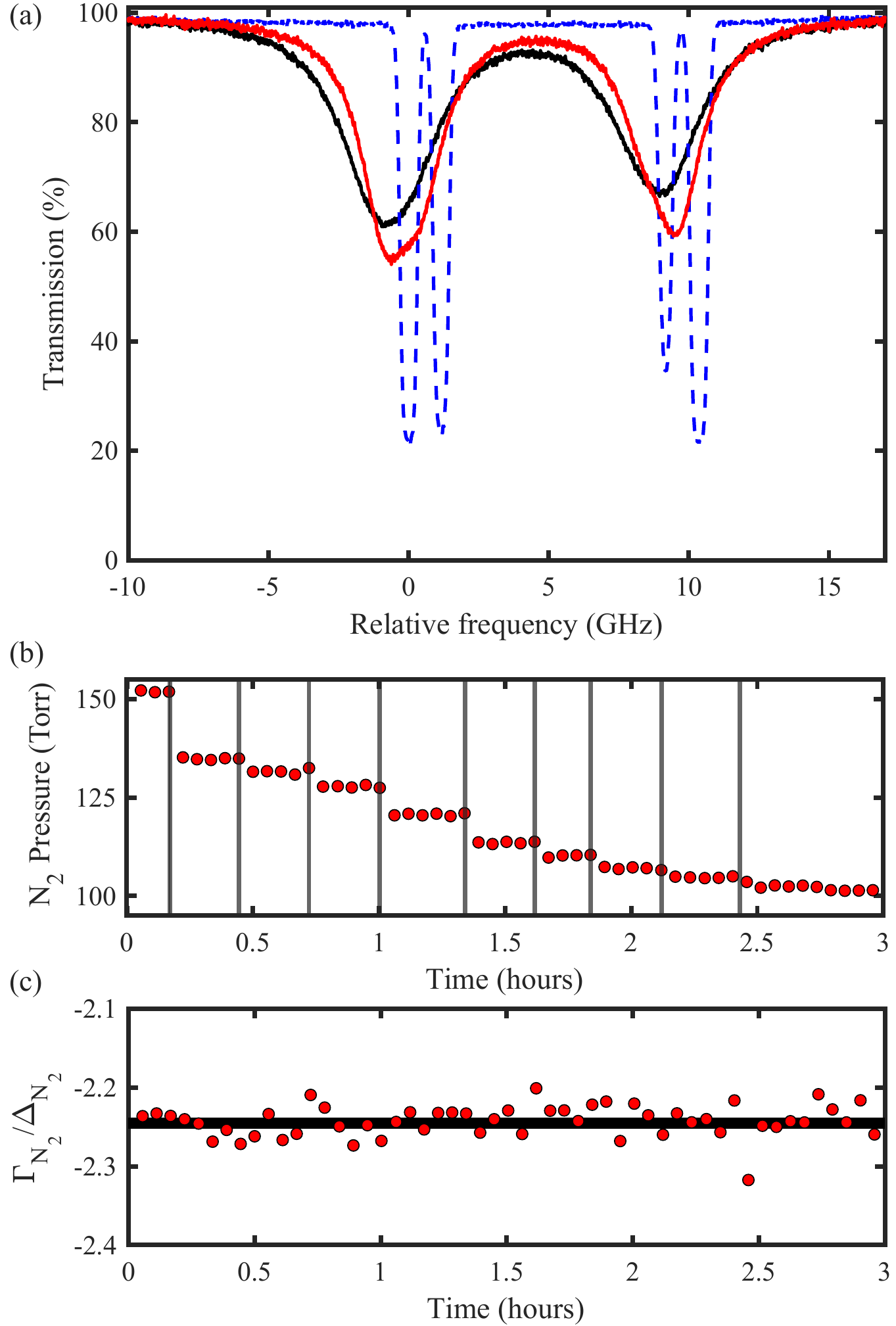}
\caption{(a): Measured spectroscopy of the Cs $D_1$ line for N$_2$ pressures of 150$\,$Torr (black), 100$\,$Torr (red) and 100$\,$mTorr (dashed blue). (b): Time sequence of the inferred nitrogen pressure as it is decreased with vertical black lines indicating when the 1070$\,$nm laser is on. (c): Ratio of the measured linewidth broadening and frequency shift as a function of time shown in red with the expected value in black.}
\label{fig2}
\end{figure}

Since the Cs in the MEMS cell is derived from the pill source, there is no initial signal until the pill source has been activated. A fibre laser at 1070$\,$nm is used for pill heating, and focused to fill the surface area of the $\approx$1$\,$mm diameter pill. The Cs source is an alkali-pill composed of Cs-chromate mixed with a non-evaporable getter (NEG) compound \cite{boudotpill,gorecki}. Recent work has shown that the heating of NEGs can be used for the passive pumping of vacuum cells by removing gases capable of diffusing into the bulk and undergoing irreversible chemical reactions with the Al and Zr mixture of the NEG \cite{mcgilliganreview, boudotmcgilligan,burrow2021,hothlittle,scherer}. Among the getterable elements, the nitrogen content within a vapor cell has been shown to be getterable \cite{mitchellneg, pillN2}. To initiate an alkali density, the pill was heated by the laser for 40$\,$s with 3$\,$W to release the Cs. After an hour of heating the cell at 100$\,\degree$C to encourage alkali diffusion through the meandering channel \cite{dyer2022}, a saturated Cs vapor was measured with the broadening and shift indicating a N$_2$ pressure of $\approx$150$\,$Torr at 60$\,^{\circ}$C. The measured initial pressure is lower than the expected pressure of 177$\,$Torr, inferred from the ideal gas law. This discrepancy could be attributed to N$_2$ consumption into the pill during the first heating pulse.

Following the release of Cs, the pill was re-heated with 3$\,$W bursts of varied duration. This level of laser power was selected as it provided a temporal response from the pill that enabled fine control of the buffer gas pressure. Spectra taken for cells tuned between 150$\,$Torr and $\leq$100$\,$mTorr are shown in Fig.~\ref{fig2} (a). During the gettering process, the N$_2$ pressure is inferred from both the linewidth and collisional frequency shift, which are extracted from fitting the spectrum with Voigt profiles. The impact of N$_2$ pressure on the $D_1$ line of Cs has been previously characterised\cite{Andalkar2002}, with coefficients for the broadening and shift determined as 19.8$\,$MHz/Torr and -8.4$\,$MHz/Torr respectively for a cell at 21$\,^{\circ}$C. The inferred N$_2$ pressure during the gettering process is shown in Fig.~\ref{fig2} (b), where the pressure is shown with red data points after each subsequent activation cycle, indicated by vertical black lines. During the pressure depletion process, the ratio of the broadening and shift was found to be $\Gamma_{BG}/\Delta_{BG}\simeq$-2.24$\pm$0.02, in good agreement with the expected value for Cs $D_1$ in the presence of N$_2$ \cite{Andalkar2002}. The tracking of this ratio is provided in Fig.~\ref{fig2} (c), where the consistently measured ratio expected for a N$_2$ dominated background pressure indicates an insignificant presence of other gas species within the cell.

\begin{figure}[t!]
\centering
\includegraphics[width=\columnwidth]{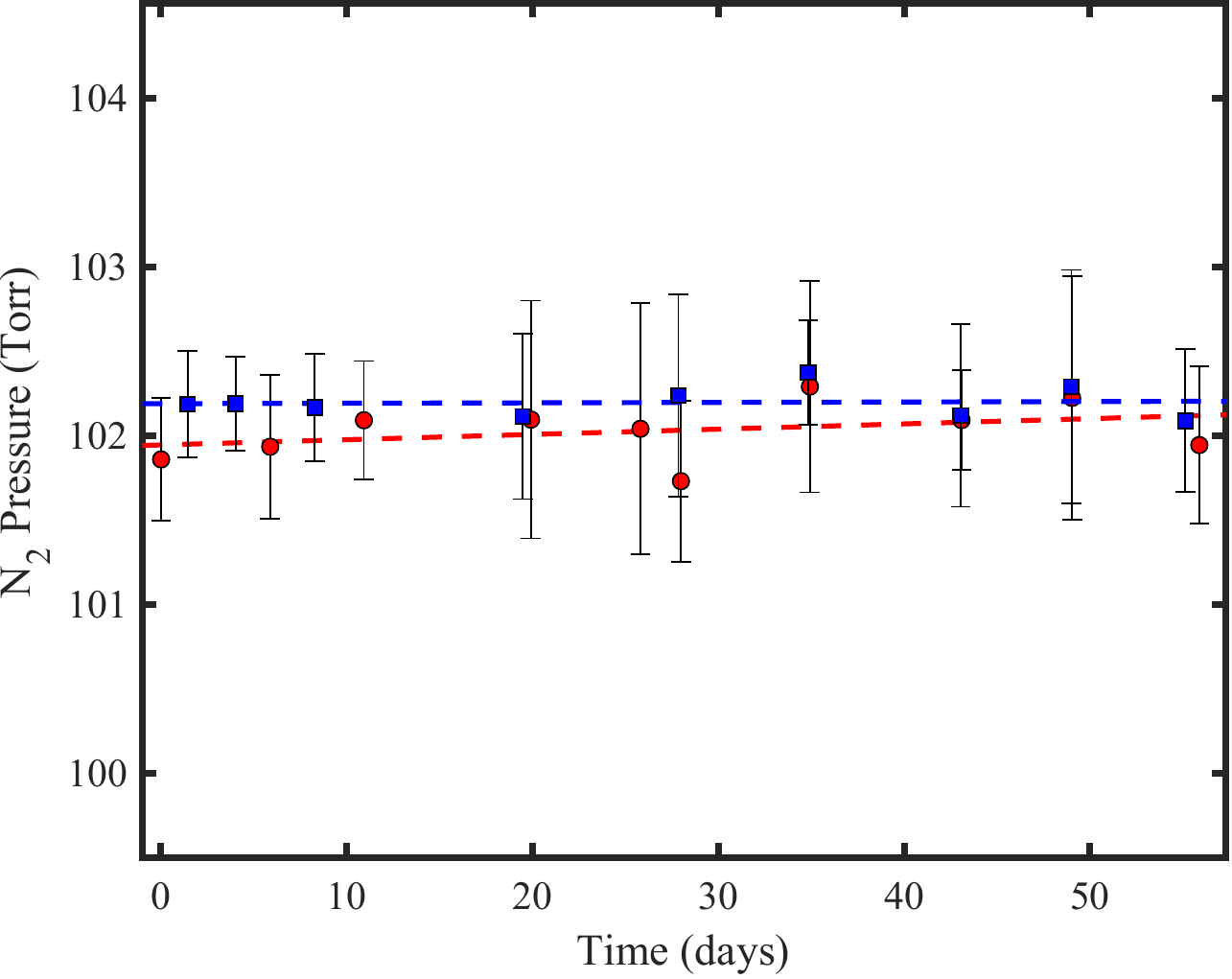}
\caption{Inferred buffer gas pressure from the shift $\Delta_{BG}$ for a cell kept at 60~$\degree$C (blue squares) and another at 120~$\degree$C (red circles).}
\label{fig3}
\end{figure}

Following the initial demonstration of controllably decreasing the N$_2$ pressure with the alkali dispensing pill, the stability of the target pressure was investigated to elucidate if the NEG compound continues to remove N$_2$ after the laser heating has been removed. For this study, two cells were tuned to P$_{N_2}\approx100\,$Torr at 60$\,^{\circ}$C while monitoring the evolution of the inner pressure environment for a period beyond 50 days. The cells were maintained at different temperatures between the subsequent measurement windows to determine if a higher cell operational temperature would impact the pressure stability, potentially occurring from an increased N$_2$ diffusion rate into the getter material at higher temperatures. Following the heating periods, the cell pressure was then measured again at an operational temperature of 60$\,^{\circ}$C. The results of this study are shown in Fig.~\ref{fig3} in blue squares and red circles for maintained temperatures of 60$^{\circ}$C and 120$^{\circ}$C respectively. Over the course of the measurement, the N$_2$ pressure remained constant around 102$\,$Torr for the measurement temperature, indicating that the gettering compound does not significantly remove content from the vapor in this cell operational temperature range. A linear fit to the cell pressure as a function of time in the cells maintained at 60$^{\circ}$C and 120$^{\circ}$C provide drifts of $\approx\,-0.3\,$mTorr/day and $\approx\,3\,$mTorr/day respectively, consistent with no change of buffer gas pressure.

\begin{figure}[t!]
\centering
\includegraphics[width=\columnwidth]{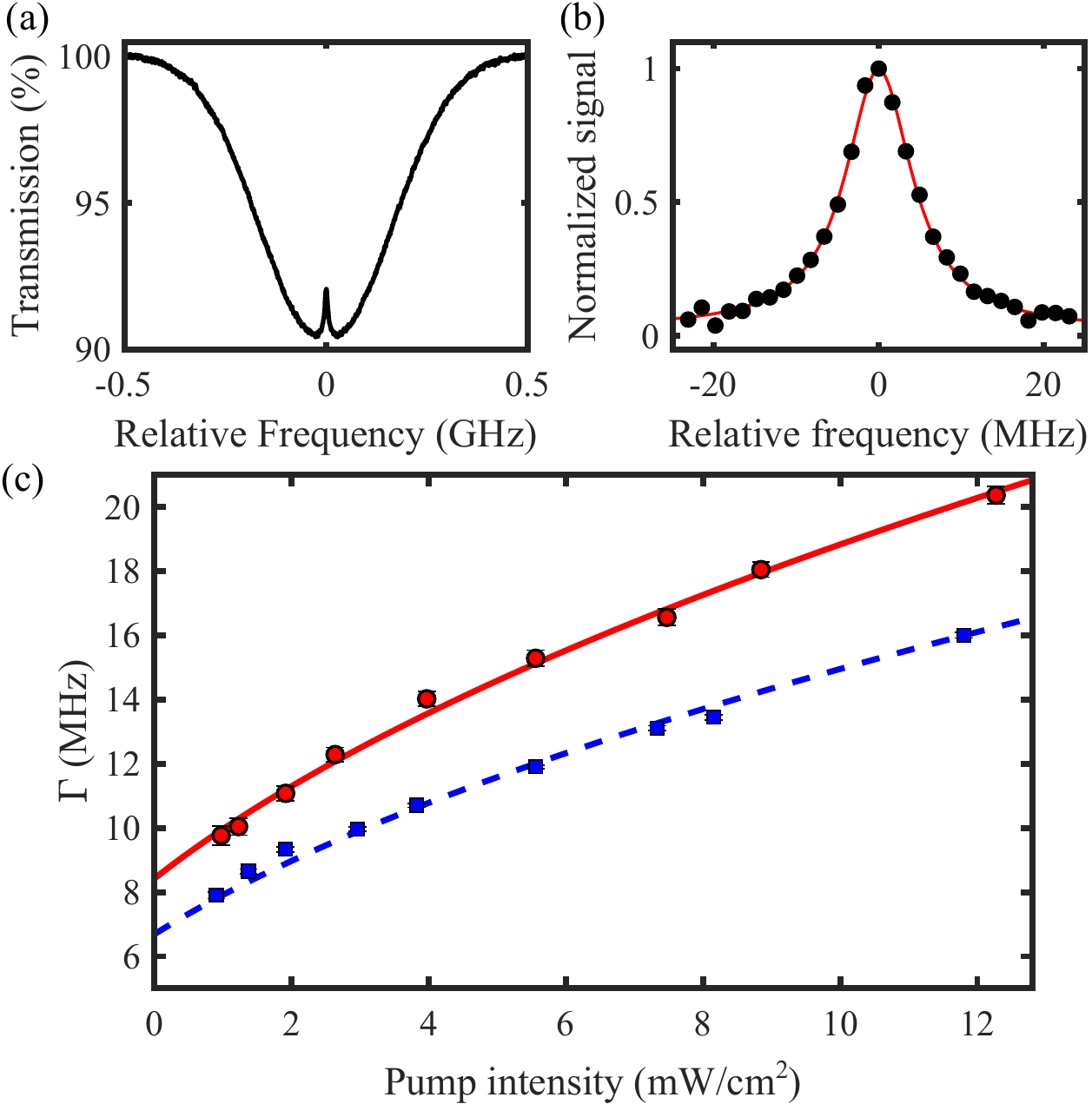}
\caption{(a): Sub-Doppler spectroscopy of the $F=4 \rightarrow F'=3$ transition (unaveraged) with pump intensity 1.2$\,$mW/cm$^2$ and vapor cell at room temperature. (b): Spectra (black circles) and Lorentzian fit (red line) of the sub-Doppler resonance extracted from (a). (c): The sub-Doppler linewidth as a function of pump intensity for the MEMS cell (red circles) and glass-blown cell (blue squares) with fit lines of the form $\Gamma=\Gamma_0\sqrt{1+I/I_{sat}}$  shown in solid red and dashed blue lines respectively.}
\label{fig4}
\end{figure}

Finally, to evaluate the presence of additional broadening mechanisms, a vapor cell was exposed to multiple laser pulses in an attempt to remove all N2 background gas. The previous apparatus was modified to allow for Doppler-free counter-propagating pump-probe spectroscopy of the $F=4\rightarrow F'=3$ transition, as shown in Fig.~\ref{fig4} (a). The linewidth of the sub-Doppler resonance is monitored for various pump beam intensities to extrapolate back to a zero-intensity linewidth. During this measurement the pump and probe beams had a waist of 0.9$\,$mm (1/e$^2$ radius) at the vapor cell and the probe beam intensity is held constant at 0.1$\,$mW/cm$^2$, to provide a sufficient signal-to-noise ratio for fitting. A typical Lorentzian fit to the sub-Doppler feature is shown in Fig.~\ref{fig4} (b). The power broadened linewidth is expected to follow the form $\Gamma=\Gamma_0\sqrt{1+I/I_{sat}}$, where $\Gamma_0=$ $2\pi\times4.5$$\,$MHz and $I_{sat}$=2.5$\,$mW/cm$^2$ \cite{SteckCS}. Here a zero-intensity linewidth of $\Gamma=$8.4$\pm0.5\,$MHz was obtained as demonstrated in Fig.~\ref{fig4} (c). To eliminate contributions from systematic broadening mechanisms such as residual Doppler broadening, laser linewidth and transit-time broadening the measurement was repeated with a glass-blown vapor cell which yielded a zero-intensity linewidth of $\Gamma=$7.0$\pm0.4\,$MHz. By comparing the zero-intensity linewidths a residual broadening of $\approx$1.4$\pm$0.6$\,$MHz, is obtained for the depleted MEMS vapor cell. If we attribute this broadening to remaining nitrogen within the cell, this would place an upper bound on the N$_2$ pressure of 100$\,$mTorr at 21$^{\circ}$C. In the future we will investigate alternative spectroscopic techniques such as two-photon spectroscopy \cite{newman2021highperformance,Georgiades:94} to better estimate the residual pressure within the cell.

In conclusion, we have demonstrated a simple approach to buffer gas pressure tuning in micro-machined vapor cells. This approach greatly simplifies current methods for in-situ tuning of the cell pressure ratio with a minimal footprint on the cell dimensions. The demonstrated tunability and expected longevity from this technique could prove advantageous to the fabrication of buffer gas based atomic sensors, such as clocks and optically pumped magnetometers, where such a method could be used for real-time fine-tuning of the sensor sensitivity. Additionally, this approach could greatly reduce manufacturing costs by enabling an array of cell target pressures to be extracted from a single wafer with a back-filled buffer gas pressure.

\begin{acknowledgments}
The authors would like to thank Dr. Jonathan Pritchard for his work on the optical absorption spectroscopy setup and Dr. Terry Dyer for his assistance and training with wafer dicing. Additionally, we would like to thank Dr. Aidan Arnold for useful conversations. The authors acknowledge funding from Defence Security and Technology Laboratory. The authors would like to acknowledge support from the INMAQS collaboration (EP/W026929/1). J. P. M gratefully acknowledges funding from a Royal Academy of Engineering Research Fellowship.
\end{acknowledgments}

\bibliography{aipsamp}

\end{document}